\documentclass[conference]{IEEEtran}
\IEEEoverridecommandlockouts
\usepackage{cite}
\usepackage{amsmath,amssymb,amsfonts}
\usepackage{algorithmic}
\usepackage{graphicx}
\usepackage{textcomp}
\usepackage{xcolor}

\usepackage{url} 

\usepackage[T1]{fontenc}
\usepackage{lmodern}
\usepackage[whole]{bxcjkjatype}

\def\BibTeX{{\rm B\kern-.05em{\sc i\kern-.025em b}\kern-.08em
    T\kern-.1667em\lower.7ex\hbox{E}\kern-.125emX}}
\begin{document}

\title{Analysis of Psychographic Indicators via LIWC and Their Correlation with CTR for Instagram Ads}

\author{\IEEEauthorblockN{Kenjiro Inoue}
\IEEEauthorblockA{\textit{Graduate School of Business Sciences} \\
\textit{University of Tsukuba}\\
Tokyo, Japan \\
s2240056@s.tsukuba.ac.jp}
\and
\IEEEauthorblockN{Mitsuo Yoshida}
\IEEEauthorblockA{\textit{Institute of Business Sciences} \\
\textit{University of Tsukuba}\\
Tokyo, Japan \\
mitsuo@gssm.otsuka.tsukuba.ac.jp}
}

\maketitle

\begin{abstract}
The online advertising industry continues to grow and accounts for over 40\% of global advertising spending.
Online display advertising consists of images and text, and advertisers maximize sales revenue by contacting consumers through advertisements and encouraging them to make purchases.
In today's society, where products are becoming more homogenized and needs are diversifying, appealing to consumer psychology through advertisements is becoming increasingly important.
However, it is not sufficiently clear what kind of appeal influences consumer psychology.
In this study, we quantified the appeal of the text in advertisements for health products and cosmetics, which were actually delivered in Instagram advertisements (one of display advertisements), by applying linguistic inquiry and word count (LIWC).
The correlation between click-through rate (CTR) and the text was analyzed. The results showed that negative appeals that arouse consumer anxiety and a sense of crisis were related to CTR.
\end{abstract}

\begin{IEEEkeywords}
Online advertising, LIWC, Instagram, text mining, CTR
\end{IEEEkeywords}

\section{Introduction}
The online advertising industry continues to grow and accounts for over 40\% of global advertising spending.\footnote{\url{https://www.marketing-interactive.com/online-advertising-to-account-for-44-6-of-global-ad-spend}}
Online advertising consists mainly of three formats: ``display advertising,'' which involves placing images and text in ad spaces on websites and apps, ``search-linked advertising,'' which displays ads on search result pages based on specific keywords entered in search engines, and ``video advertising,'' which uses video file formats (audiovisual content).

With the growth of the online advertising market, related research has also increased, particularly in the prediction of CTR (Click-through Rate).
CTR refers to the ratio of the number of times an ad is clicked to the number of times it is displayed (impressions), indicating that a higher CTR signifies ads that attract more consumer interest and engagement.
Since logistic regression is used to report CTR prediction~\cite{Richardson2007}, one of the multivariate analysis methods, various analytical models such as decision trees and linear regression analysis have also been reported to be useful for CTR prediction~\cite{Effendi2017,Trofimov2012}.
Furthermore, many of the reported CTR prediction models since the late 2010s have utilized deep learning.
One representative deep learning model is DeepFM (deep factorization-machine)~\cite{Guo2017}, proposed in 2017.
DeepFM is an improvement of the Wide \& Deep model proposed by Cheng et al.~\cite{Cheng2016}, capable of learning low-dimensional features (categorical variables), such as URL and landing page characteristics, and high-dimensional features, such as images, while considering the interactions between variables.
An even more accurate model, xDeepFM~\cite{Lian2018}, which partially improves the layer structure of DeepFM, has been reported, and its API has already been released.
Thus, research in online advertising is expected to continue accelerating in the future.

In the field of psychology, traditional approaches, such as experiments and surveys, have been primarily used to analyze individuals' behavioral tendencies and values.
However, with the development of the advanced information society, the utilization of big data, which allows us to grasp people's social activities through their online behavioral histories, is becoming increasingly important.
Among them, the analytical approach related to language content, which infers psychological tendencies from digitally captured text data, has gained significant global attention.
Specifically, by conducting dictionary-based analysis on text data obtained from social media platforms like Twitter, it becomes possible to quantify consumer psychology and behavior, thereby gaining insights.
Representative dictionary-based analysis methods include ML-Ask~\cite{Ptaszynski2009,Ptaszynski2017}, which measures ten emotional components, joy, anger, anticipation, sadness, liking, fear, trust, disgust, surprise, and shame, based on text; LIWC2015~\cite{Pennebaker2015}, which assigns and measures linguistic and psychological categories to individual words comprehensively; and moral foundations dictionary (MFD)~\cite{Graham2009}, which measures morals associated with each text based on the moral foundation theory~\cite{Graham2013}.
Psychological research using text data, such as the application of MFD to Twitter tweets~\cite{Matsuo2019}, has been actively conducted.

In online advertising too, it has become possible to infer the linguistic content of appeals and link it to advertising performance metrics such as CTR and CVR (conversion rate; the proportion of website visitors who completed a desired action, such as a purchase or inquiry).
However, while research on advertising performance prediction is thriving, the relationship between the linguistic content of appeals and advertising performance metrics remains insufficiently understood.

In this study, we conducted a preliminary investigation focusing on the Japanese advertising market.
Specifically, we analyzed advertisements for health products and cosmetics that were composed in the Japanese language and delivered through Instagram, which is one of the major platforms for online display advertising.
We quantified the appeals of the actual delivered advertisement texts, using the Japanese version of LIWC (linguistic inquiry and word count)~\cite{Igarashi2022} and analyzed their correlation with CTR.
As a result, it was found that negative appeals that evoke consumer anxiety or crisis were related to CTR.
Additionally, when the advertisement text included price information, there was a negative correlation between CTR and health products, while a positive correlation was observed for cosmetics.

\section{Data and Basic Statistics}

\begin{table*}[t]
  \caption{Summary of data: ``Avg.CTR'' is the average CTR per ad unit. Avg.CTR for health products (0.75\%) is higher than avg.CTR for Cosmetics (0.61\%).}
  \label{tb:Data}
  \centering
\begin{tabular}{ll|rr}
\hline
\hline
\multicolumn{1}{l}{Features} & \multicolumn{1}{l|}{Details} & \multicolumn{2}{c}{Product category} \\
\multicolumn{1}{c}{} & \multicolumn{1}{c|}{} & \multicolumn{1}{c}{Health products} & \multicolumn{1}{c}{Cosmetics} \\ \hline
Product count & The number of products that ran Instagram ad campaigns. & 48 & 45 \\
Ad count & The number of unique ads. & 3,555 & 3,270 \\
Avg.CTR & Average CTR per ad unit. & 0.75\% & 0.61\% \\
\hline
\end{tabular}
\end{table*}

\begin{table*}[t]
  \caption{Character counts and proportions relative to the total character count: In both categories, symbols account for about 10\% of the total, because they contain decorative symbols to attract attention.}
  \label{tb:Characters}
  \centering
\begin{tabular}{l|rr|rr}
\hline
\hline
 & \multicolumn{2}{c|}{Health products} & \multicolumn{2}{c}{Cosmetics} \\
 & \multicolumn{1}{c}{Main text} & \multicolumn{1}{c|}{In-image text} & \multicolumn{1}{c}{Main text} & \multicolumn{1}{c}{In-image text} \\ \hline
Numbers & 2.7 (\phantom{0}3.4\%) & 15.4 (\phantom{0}9.5\%) & 3.1 (\phantom{0}3.2\%) & 6.5 (\phantom{0}4.4\%) \\
Alphabetic characters & 1.2 (\phantom{0}1.4\%) & 27.6 (14.0\%) & 4.1 (\phantom{0}4.7\%) & 72.7 (45.3\%) \\
Katakana & 17.4 (24.8\%) & 31.8 (14.4\%) & 16.8 (18.3\%) & 13.5 (10.0\%) \\
Hiragana & 20.4 (28.2\%) & 33.1 (23.7\%) & 22.8 (25.4\%) & 16.9 (12.9\%) \\
Kanji & 23.3 (32.2\%) & 49.7 (27.8\%) & 29.1 (33.8\%) & 20.7 (16.8\%) \\
Symbols & 6.8 (\phantom{0}9.2\%) & 12.1 (\phantom{0}6.9\%) & 9.6 (10.7\%) & 7.3 (\phantom{0}4.8\%) \\
Emoji & 0.8 (\phantom{0}0.8\%) & 0 (\phantom{0}0.0\%) & 3.6 (\phantom{0}3.9\%) & 0 (\phantom{0}0.0\%) \\
\hline
Character count & 72.5 & 175.8 & 89.2 & 146.5 \\
\hline
\end{tabular}
\end{table*}

\subsection{Instagram ads}

\begin{figure}[tp]
    \centering
    \includegraphics[width=0.98\linewidth]{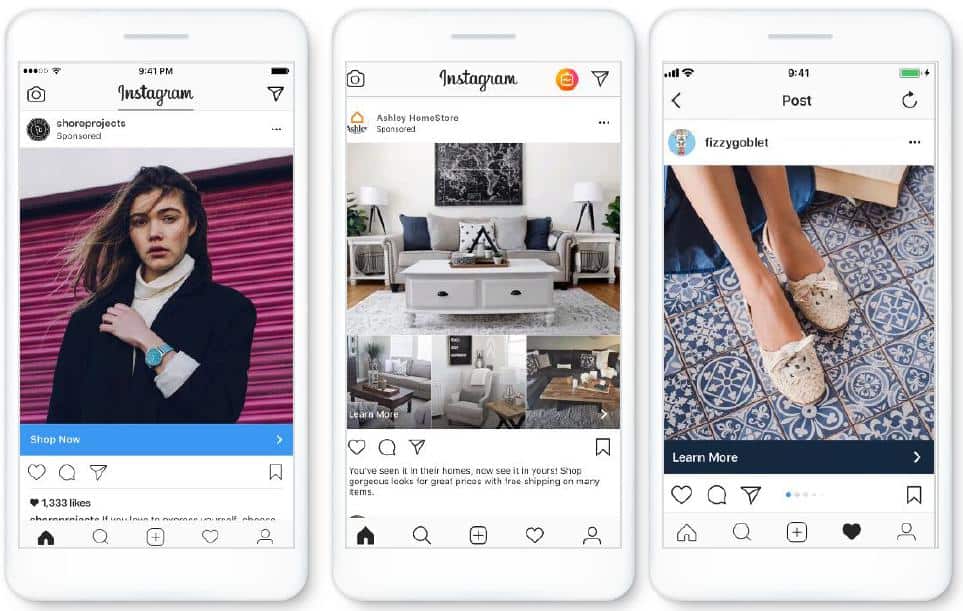}
    \caption{Instagram ads: Users who use Instagram have a high awareness of images and videos, and it is desirable to have advertisements that do not create a sense of discomfort.}
    \label{fig:advertisement}
\end{figure}

Instagram\footnote{\url{https://www.instagram.com/}} has 1 billion monthly active accounts worldwide, including 33 million in Japan.
Instagram advertising, one of the advertising options provided by Meta's Facebook Ads, allows advertisers to deliver ads on the Instagram platform.
Fig.~\ref{fig:advertisement} is an example of Instagram advertising.
Detailed targeting options are available, including user demographics, such as age, gender, occupation, and interests.
Instagram is a social media platform focused on posting and viewing photos and videos, attracting users who are highly conscious of visual content.
Therefore, in Instagram advertising, creating compelling creatives (including ad texts and images) that capture users' attention without causing any sense of discomfort is crucial.

\subsection{Data}

In this survey, we focused on the Japanese advertising market and specifically analyzed the categories of health products and cosmetics. We used the performance data of Instagram ads delivered between January 1, 2020, and December 31, 2022.
The overview of the data used is presented in TABLE.~\ref{tb:Data}.
The ``Product count'' represents the total number of products advertised on Instagram, with 48 products in the health products category and 45 products in the cosmetics category.
The ``Ad count'' represents the total number of unique ads consisting of combinations of images and main texts delivered for each product.
We collected 3,555 ad data for health products and 3,270 ad data for cosmetics.
Please note that, to ensure the effectiveness of ad-level CTR, ads with an impression count below 10,000 were excluded from this survey.
The ``Ave.CTR'' calculates the average CTR at the ad level. The ad-level CTR is derived from the ratio of the number of clicks to the impression count.

\begin{figure}[tp]
    \centering
    \includegraphics[width=0.98\linewidth]{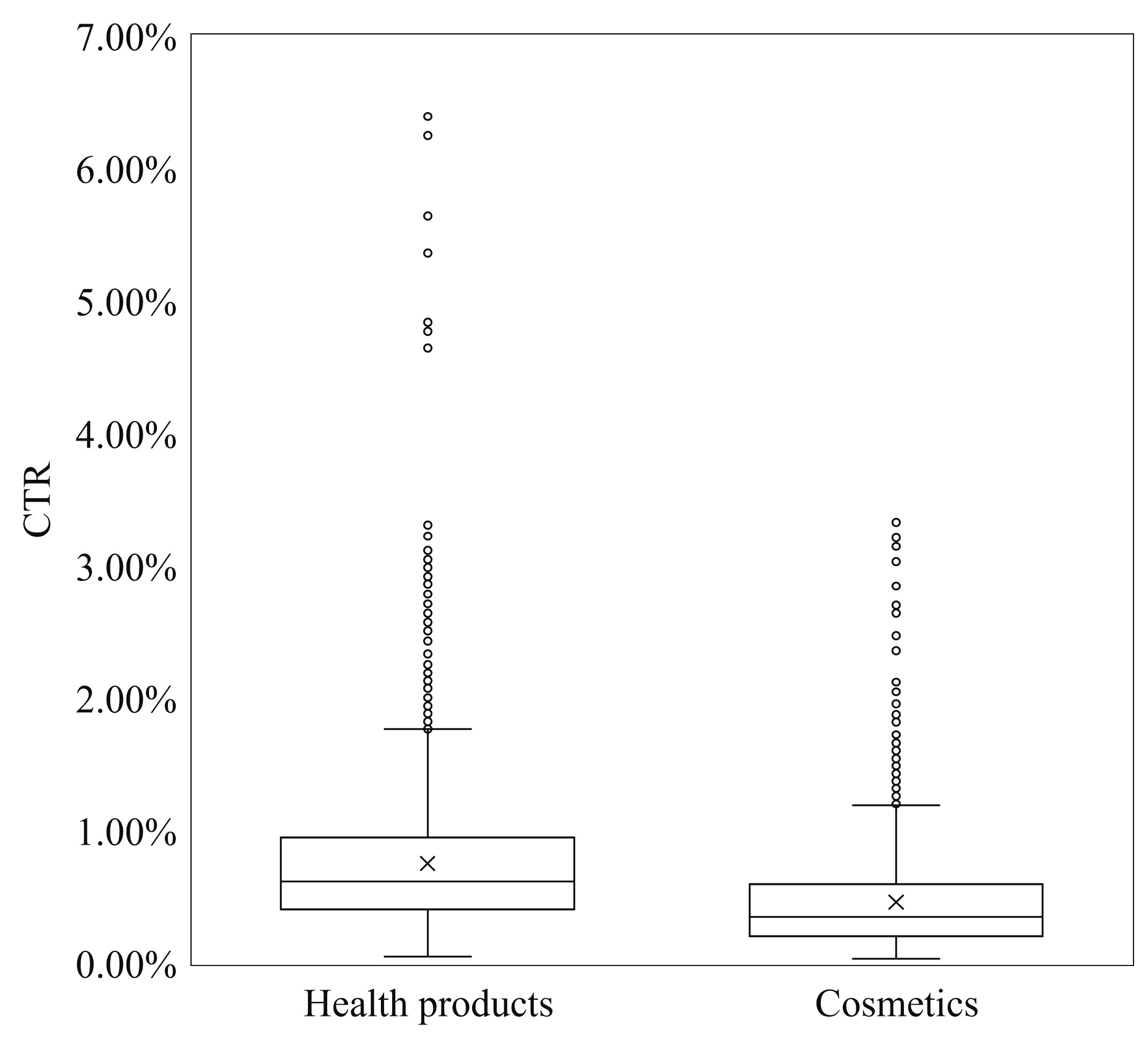}
    \caption{Product category CTR box plot: Compared to health products, cosmetics have a smaller distribution of CTR. Cosmetics primarily promise benefits related to beautiful and healthy skin maintenance, resulting in relatively consistent effects across different products.This leads to increased competition and a tendency for a smaller CTR distribution.}
    \label{fig:Product Category CTR Box Plot}
\end{figure}

The box plot visualizing the CTR distribution for health products and cosmetics is shown in Fig.~\ref{fig:Product Category CTR Box Plot}.
The ave.CTR for health products is 0.75\%, which is higher compared to 0.61\% for cosmetics, indicating a larger CTR distribution for health products.
Cosmetics primarily promise benefits related to beautiful and healthy skin maintenance, resulting in relatively consistent effects across different products.
This leads to increased competition and a tendency for a smaller CTR distribution.
On the other hand, health products encompass a wide range of food products that claim to be beneficial for health.
The expected effects may vary, including health maintenance, fat suppression, and prevention or improvement of high blood pressure.
This variability suggests a larger CTR distribution compared to cosmetics.

\subsection{Preparation of text data}

From the advertisements, two types of text can be extracted: ``main text,'' which is mandatory and displayed alongside the image when delivering Instagram ads, and ``in-image text,'' which is text displayed within the image.
The main text is limited to 125 characters as per Instagram ad guidelines, while there is no character limit to the in-image text.
The in-image text was extracted using the OCR (optical character recognition) feature of the Google Vision API\footnote{\url{https://cloud.google.com/vision}}.

The extracted text data from the advertisements were categorized into ``Numbers,'' ``Alphabetic characters,'' ``Katakana,'' ``Hiragana,'' ``Kanji,'' ``Symbols,'' and ``Emojis'' for both the main text and in-image text.
The aggregated results, including the average character count and the percentage of each category in relation to the total text, are presented in TABLE.~\ref{tb:Characters}.
In the main text, the total character count is higher for cosmetics than for health products.
Furthermore, both categories have approximately 80\% of their text composed of Hiragana, Katakana, and Kanji, while symbols account for around 10\% of the text.
This is due to the inclusion of decorative symbols aimed at catching attention (``【お一人様一回限り】'' (One time only per person) and ``＼先着15,000名様限定！／'' (Limited to the first 15,000 applicants!)).
In cosmetics, emojis are used for the same purpose, as they are targeting mainly female audiences, resulting in a higher proportion compared to health products.
The higher usage of alphabetic characters in cosmetics suggests the presence of product names written in English within the main text.

In the case of in-image text, since OCR was used to extract the text, it was unable to detect unreadable characters.
For example, emojis used within the images were not recognized as text.
Moreover, the recognition of the text displayed in product images resulted in an unusually high proportion of alphabetic characters, particularly in cosmetics, accounting for 45.3\%. This highlights the need for preprocessing steps such as stop-word removal.
On the other hand, when examining the character count for Hiragana, Katakana, Kanji, and symbols, health products had a higher count compared to cosmetics, showing a contrasting trend to the main text results.
This suggests that, in the case of health products, information related to the products and appeals to consumers are more likely to be expressed within the images rather than in the main text.

\begin{figure}[tp]
    \centering
    \includegraphics[width=0.98\linewidth]{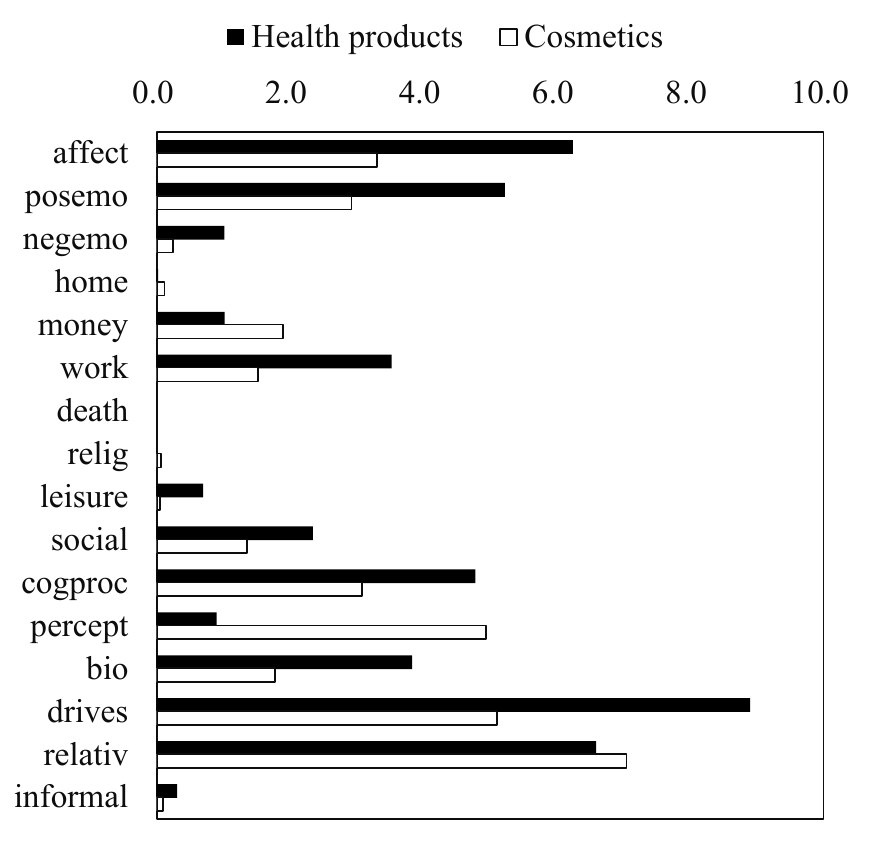}
    \caption{Distribution of mean values by psychological categories (main text): In health products, the occurrence rate of ``cause'' is high, and it contains words such as ``ideal,'' ``insufficient,'' ``countermeasure,'' and ``method'' that stimulate triggers.}
    \label{tb:fig3}
\end{figure}

\begin{figure}[t]
    \centering
    \includegraphics[width=0.98\linewidth]{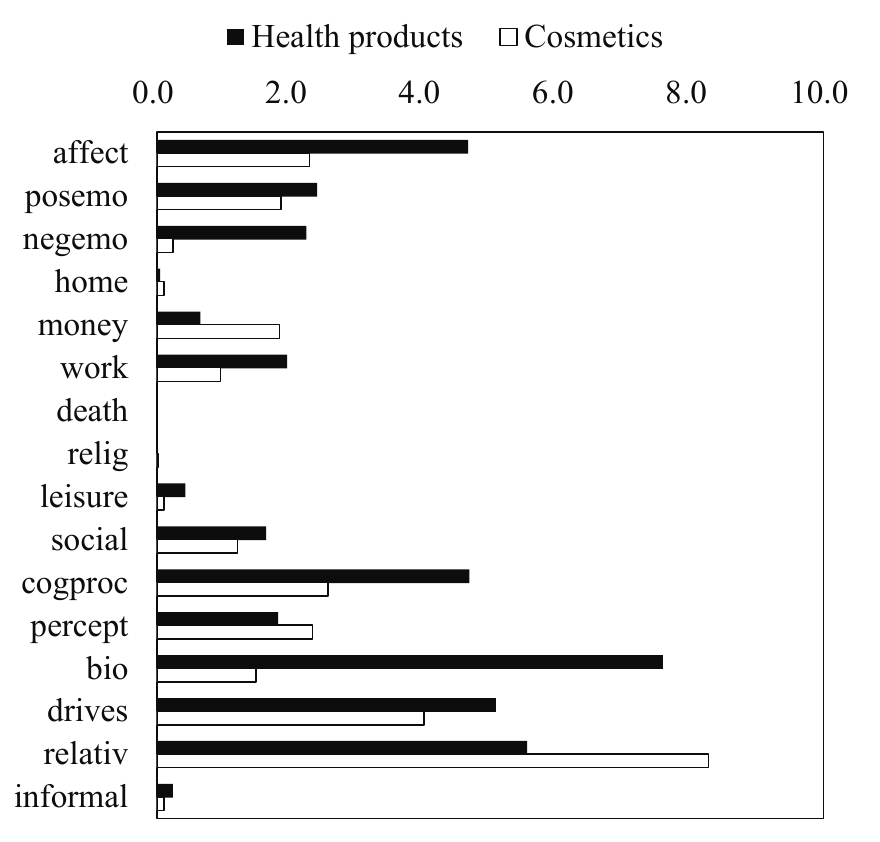}
    \caption{Distribution of mean values by psychological categories(In-image text): Cosmetics, similar to the main text, have a higher occurrence rate of ``relativi,'' and include words that indicate temporal deadlines, such as ``特別キャンペーン12月31日まで！'' (Special campaign until December 31st!).}
    \label{tb:fig4}
\end{figure}

\subsection{Application of J-LIWC2015}

The text data of the advertisements were quantified by applying the Japanese version of LIWC2015 (J-LIWC2015), developed by Igarashi et al.~\cite{Igarashi2022}.
Linguistic inquiry and word count\footnote{\url{http://www.liwc.net/}} is a standard dictionary for text analysis that assigns linguistic categories (including function words, as auxiliary verbs and conjunctions, and content words, such as nouns and verbs) and psychological process categories (including emotional processes, cognitive processes, and social processes) to commonly used words based on pre-established classifications by experts.
Prior to the analysis of the text data, the J-LIWC2015 dictionary was applied using the source code\footnote{\url{https://github.com/tasukuigarashi/j-liwc2015}} provided by Igarashi to perform the necessary tagging and calculate the occurrence ratios of words in each category present in the text.

The distribution of average values for psychological categories is shown in Fig.~\ref{tb:fig3} for main text and Fig.~\ref{tb:fig4} for in-image text.
In the main text, health products have a higher occurrence ratio in the ``Causation (cause)'' category, while cosmetics have a higher occurrence ratio in the ``Relativity (relativ)'' category.
In the in-image text, health products have a higher occurrence ratio in the ``Biological processes (bio)'' category, while cosmetics show a similar higher occurrence ratio in the ``Relativity (relativ)'' category as in the main text.
The ``Causation (cause)'' category in health products includes words, such as ``理想 (ideal),'' ``不足 (shortage),'' ``対策 (countermeasures),'' and ``方法 (methods),'' that prompt triggers, while the ``Biological processes (bio)'' category includes words related to biological aspects, such as ``疲労 (fatigue),'' ``脂肪 (body fat),'' and ``血圧 (blood pressure).''
This suggests that health products often employ advertisements with a composition where the in-image text evokes consumers' physical concerns and the main text encourages actions.
The ``Relativity (relativ)'' category pertains to actions, space, and time.
In the case of cosmetics, it includes expressions indicating temporal deadlines, such as ``特別キャンペーン12月31日まで！'' (Special campaign until December 31st!).
This suggests that cosmetics often utilize advertisements with a composition that creates a sense of something unique through campaigns and prompts actions by providing a time limit.

\section{Analysis and Results}

In this section, we analyze and discuss the correlation between the main text and in-image text of advertisements and the psychological processes measured by J-LIWC2015 as well as their correlation with CTR for both the health products and cosmetics categories.
The correlation coefficient $\rho$ is described as follows:
\begin{equation}
\rho = \frac{\sum{(X_i - \bar{X})(Y_i - \bar{Y})}}{\sqrt{\sum{(X_i - \bar{X})^2} \sum{(Y_i - \bar{Y})^2}}}
\end{equation}
where $X_{i}$ is the CTR, and $Y_{i}$ is the number of occurrences of the word for each category of psychological processes in the text.
Each value is in units of unique ads.
The size of the array is the same as Ad count, which is $|X|=3555$ for health products.

\subsection{Health products}

\begin{table}[t]
  \caption{Correlation coefficients between psychological category unit values and CTR for health products: Bolded values indicate values that are relatively correlated within a psychological category.}
  \label{tb:CTR(Health products)}
  \centering
\begin{tabular}{ll|rr}
\hline
\hline
\multicolumn{1}{c}{} & \multicolumn{1}{c|}{} & \multicolumn{1}{c}{Main text} & \multicolumn{1}{c}{In-image text} \\ \hline
\multicolumn{2}{l|}{affect} & -0.029 & \textbf{0.218} \\
 & posemo & -0.130 & -0.058 \\
 & negemo & \textbf{0.221} & \textbf{0.302} \\ \hline
personal concerns & home & -0.066 & -0.024 \\
 & money & 0.056 & \textbf{-0.170} \\
 & work & -0.044 & \textbf{-0.190} \\
 & death & - & 0.006 \\
 & relig & - & -0.034 \\
 & leisure & -0.120 & -0.013 \\ \hline
\multicolumn{2}{l|}{social} & -0.033 & -0.116 \\
\multicolumn{2}{l|}{cogproc} & 0.152 & 0.143 \\
\multicolumn{2}{l|}{percept} & 0.027 & -0.099 \\
\multicolumn{2}{l|}{bio} & \textbf{0.151} & 0.022 \\
\multicolumn{2}{l|}{drives} & \textbf{-0.182} & 0.115 \\
\multicolumn{2}{l|}{relativ} & 0.013 & -0.049 \\
\multicolumn{2}{l|}{informal} & -0.148 & -0.003 \\ \hline
\end{tabular}
\end{table}

\begin{figure}[t]
    \centering
    \includegraphics[width=0.98\linewidth]{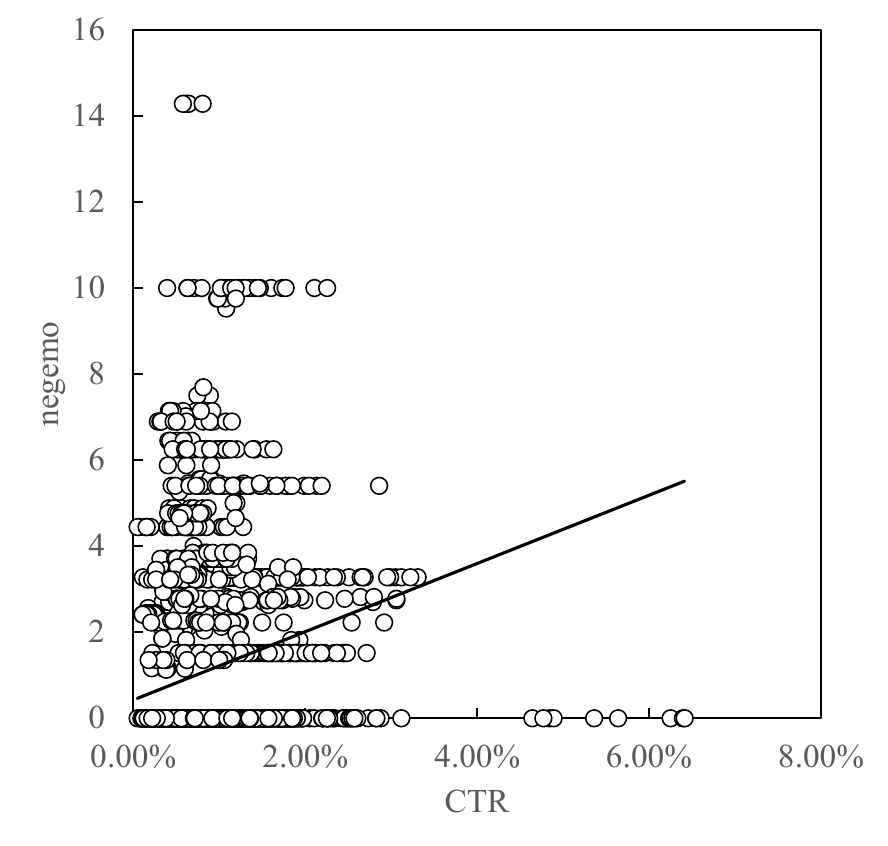}
    \caption{Health products (main text): Health products have a weak positive correlation between ``Negative emotions (negemo)'' and CTR. It is suggested that CTR increases by arousing target consumers' health consciousness through negative appeals when promoting the efficacy of the product.}
    \label{tb:fig5}
\end{figure}

\begin{figure}[t]
    \centering
    \includegraphics[width=0.98\linewidth]{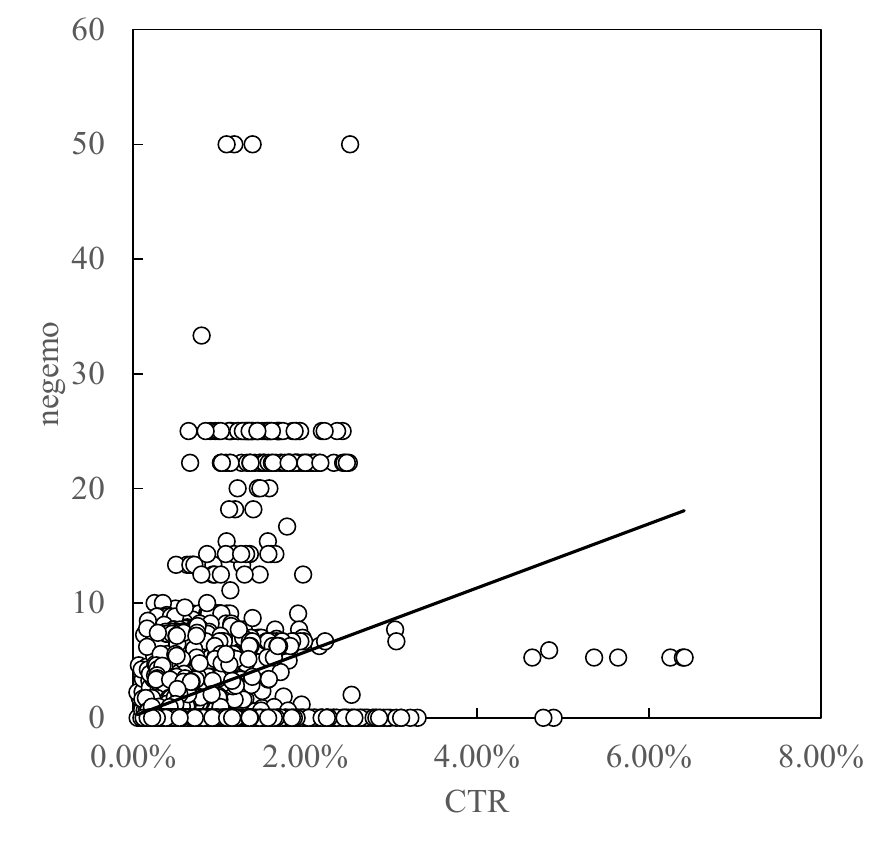}
    \caption{Health products (in-image text): As with in-image text, there is a weak positive correlation between ``Negative emotions (negemo)'' and CTR.}
    \label{tb:fig6}
\end{figure}

The correlation coefficients between the psychological process indicators and CTR for advertisements were calculated separately for the main text (main) and in-image text (image). The results, computed at the category level of psychological processes, are presented in TABLE.~\ref{tb:CTR(Health products)}.

The main text showed a weak positive correlation with ``Negative emotions (negemo)'' and ``Biological processes (bio)'' compared to other categories, while a weak negative correlation was observed with ``Drives (drives)''.
In the case of in-image text, there was a weak positive correlation with ``Affective processes (affect)'' and ``Negative emotions (negemo)'' relative to other categories, and a weak negative correlation was observed with ``personal concerns (work)'' and ``personal concerns (money)''.
Both the main text and in-image text exhibited a positive correlation with ``Negative emotions (negemo)'' among the categories of psychological processes (see Fig.~\ref{tb:fig5} and Fig.~\ref{tb:fig6}).

In the case of health products, when promoting the efficacy and benefits of the products, there is a tendency for CTR to increase by evoking a sense of sense of urgency regarding the health of the target consumers through negative appeals.
Furthermore, it is believed that including biological terms, such as ``疲労 (fatigue),'' ``脂肪 (body fat),'' and ``血圧 (blood pressure),'' when promoting the efficacy and benefits contributes to an increase in CTR.
Additionally, advertisements with a high value in the category of ``personal concerns (money)'' include expressions related to product prices, such as ``お試し価格500円(税込)'' (Trial price of 500 yen (tax included)).
Users may recognize these advertisements and hesitate to click on them.
Based on these results, it is suggested that for health products, it is effective to avoid including expressions related to product prices to prevent users from perceiving the advertisement, and instead, to evoke a sense of urgency regarding the health consciousness of the target consumers through negative appeals, which can lead to an increase in CTR.

\subsection{Cosmetics}

\begin{table}[t]
  \caption{Correlation coefficients between psychological category unit values and CTR for cosmetics: Bolded values indicate values that are relatively correlated within a psychological category.}
  \label{tb:CTR(Cosmetics)}
  \centering
\begin{tabular}{ll|rr}
\hline
\hline
 &  & Main text  & In-image text \\ \hline
\multicolumn{2}{l|}{affect} & \textbf{0.226} & -0.025 \\
 & posemo & 0.138 & -0.073 \\
 & negemo & \textbf{0.254} & 0.028 \\ \hline
personal concerns & home & 0.060 & 0.114 \\
 & money & \textbf{0.259} & -0.076 \\
 & work & -0.024 & 0.127 \\
 & death & - & - \\
 & relig & 0.126 & -0.024 \\
 & leisure & 0.078 & -0.022 \\ \hline
\multicolumn{2}{l|}{social} & -0.009 & 0.002 \\
\multicolumn{2}{l|}{cogproc} & 0.164 & 0.085 \\
\multicolumn{2}{l|}{bio} & \textbf{-0.308} & 0.020 \\
\multicolumn{2}{l|}{percept} & 0.028 & \textbf{0.259} \\
\multicolumn{2}{l|}{drives} & 0.194 & -0.028 \\
\multicolumn{2}{l|}{relativ} & 0.106 & \textbf{-0.192} \\
\multicolumn{2}{l|}{informal} & 0.094 & -0.022 \\ \hline
\end{tabular}
\end{table}

\begin{figure}[t]
    \centering
    \includegraphics[width=0.98\linewidth]{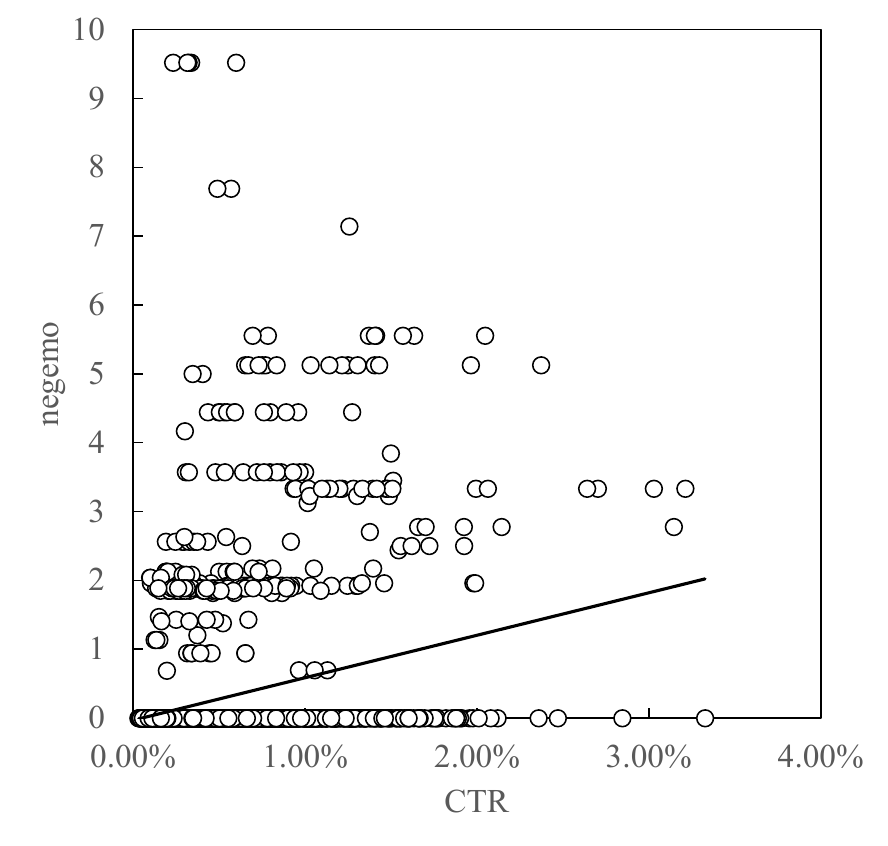}
    \caption{Cosmetics (main text): Similar to health products, cosmetics also show a positive correlation with ``Negative emotions (negemo)'', suggesting that arousing a sense of urgency related to aging and skin health through negative appeals contributes to an increase in CTR.}
    \label{tb:fig7}
\end{figure}

\begin{figure}[t]
    \centering
    \includegraphics[width=0.98\linewidth]{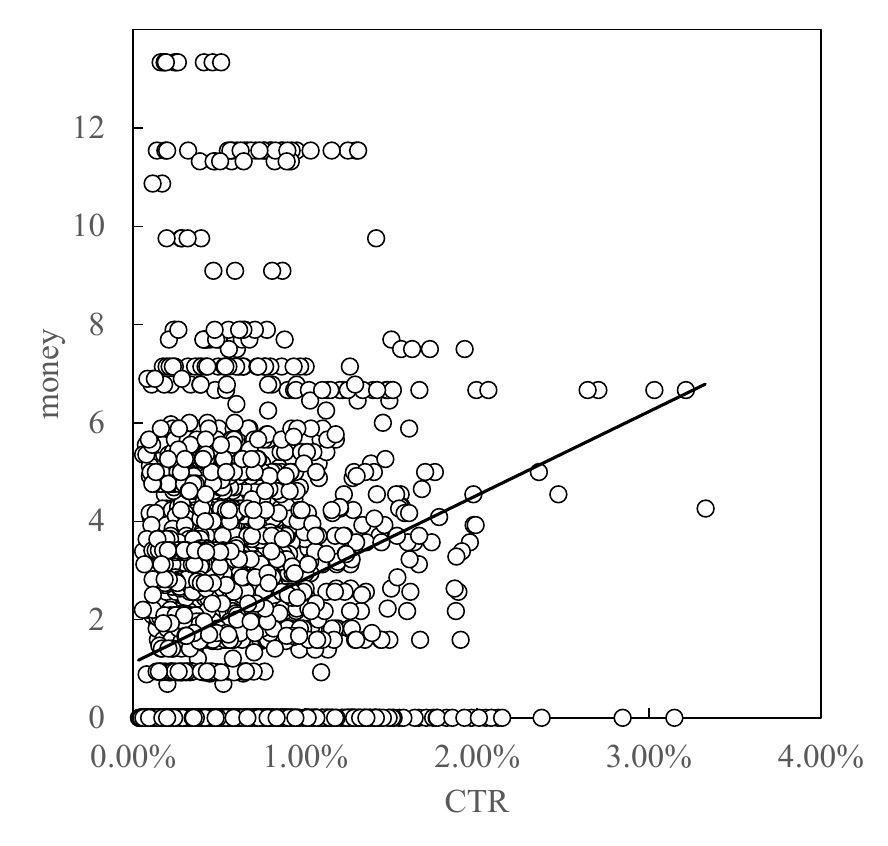}
    \caption{Cosmetics (in-image text): Cosmetics exhibit a positive correlation with ``personal concern (money)'', which contrasts with the results of health products.}
    \label{tb:fig8}
\end{figure}

Similarly to health products, cosmetics also show the correlation coefficients in TABLE.~\ref{tb:CTR(Cosmetics)}.

In the main text, there is a relatively weak positive correlation with ``Affective processes (affect),'' ``Negative emotions (negemo),'' and ``personal concern (money)'' compared to other categories, and a weak negative correlation with ``Perceptual processes (percept).''
Perceptual processes involve the senses, experiences, and expressions related to skin perception, such as ``明るい (bright),'' ``ツルツル (smooth),'' and ``つやつや (glossy),'' may potentially decrease CTR.
In the in-image text, there is a relatively weak positive correlation with ``Biological processes (bio)'' and a weak negative correlation with ``Relativity (relativ)'' compared to other categories. 
The ``Biological processes (bio)'' category includes aspects related to the body, such as ``肌 (skin),'' and explicitly mentioning direct effects to and benefits for the skin may be a factor contributing to CTR increase.
The ``Relativity (relativ)'' category refers to indicators related to actions, space, and time, and expressions indicating temporal limitations like ``特別キャンペーン12月31日まで！'' (Special campaign until December 31st!) may potentially decrease CTR.

Similar to health products, cosmetics also show a positive correlation with ``Negative emotions (negemo)'' (Fig.~\ref{tb:fig7}), suggesting that arousing a sense of urgency related to aging and skin health through negative appeals contributes to an increase in CTR.
On the other hand, cosmetics exhibit a positive correlation with ``personal concern (money)'' (Fig.~\ref{tb:fig8}), which contrasts with the results of health products.
For cosmetics, indicating the product price is suggested to contribute to an increase in CTR.

\section{Conclusion}

In this study, we conducted a foundational investigation using the Japanese version of LIWC to quantitatively measure psychological indicators from the texts of actual advertising creatives delivered on Instagram. Our aim was to reveal the relationship with these indicators and CTR through correlation analysis.
The results showed that both health products and cosmetics had a significant relationship between CTR and negative appeals that evoke consumer anxiety and crisis awareness.
Furthermore, we observed a difference in the representation of product prices. While health products exhibited a negative correlation with CTR, cosmetics showed a positive correlation.

\bibliographystyle{IEEEtran}
\bibliography{reference}

\end{document}